# The anisotropic photorefractive effect in lithium sulfo-phosphate glass system doped with nickel ions

A.Siva Sesha Reddy [a], A.V.Kityk [b,*], J.Jedryka [b,*], N.Purnachand [c], P.Rakus [b], A.Wojciechowski [b], A.S.Andrushchak [d], V.Ravi Kumar [a,*], N.Veeraiah [a]

[a] *Department of Physics, Acharya Nagarjuna University, Nagarjuna Nagar, India*
[b] *Faculty of Electrical Engineering, Czestochow University of Technology, Czestochowa, Poland*
[c] *School of Electronics Engineering, VIT-AP University, Inavolu, 522237, AP, India*
[d] *Institute of Applied physics and Nanomaterials science, Lviv Polytechnic National University, Lviv, 79046, Ukraine*

## ABSTRACT

In this work, nonlinear optical (NLO) studies of nickel oxide doped (ranging from 0.2 to 1.0 mol %) $Li_2SO_4$–MgO–$P_2O_5$ glasses are reported. A combination of femtosecond (fs) laser, as a pumping light source and a high-accuracy polarimeter with low power probing laser, is used to investigate the light-induced optical anisotropy (OA) in these glass materials. The light-induced birefringence (LIB) exhibits slow relaxation tendency up to about 10 s suggesting on anisotropic photorefractive effect evidently dominated fast Kerr effect. This behavior is evaluated in the light of other results reported recently and is explained by the glass polymerization mechanisms. The photorefractive birefringence increases with increase of the quantity of NiO up to 0.8 mol% and it is attributed to the enhanced degree of depolymerization of the glass network due to the hike in the concentration of $Ni^{2+}$ ions that occupy octahedral ($O_h$) positions. Further increase of NiO content ( > 0.8 mol%) causes, however, a certain decrease of the photorefractive birefringence. Notable change in concentration trend is interpreted by rising of phonon losses due to increasing portion of the nickel ions occupying tetrahedral ($T_h$) sites that facilitates the polymerization of the glass network. The doped glasses with the NiO content of about 0.8 mol% may be considered as optimal in the sense of photorefractive efficiency. Relevant samples exhibit largest magnitudes of the photorefractive birefringence and appear to be favorable for potential applications.

## 1. Introduction

Since last few decades, nonlinear optical (NLO) effects in glass materials have become an important area of research both in fundamental and applied aspects due to their important uses in the fields of opto-electronics and photonics, such as e.g. nonlinear optical converters (third harmonic generation, THG), 3D photonic instruments for inte-grated optics, broad band optical amplifiers, high speed optical switches, power limiters as well as in a number of photorefractive ap-plications [1,2]. Relevant studies deal with NLO phenomena causing modification of the optical material properties by the presence of intense electromagnetic optical fields [3]. The transparent amorphous materials are optically isotropic and macroscopically centrosymmetric. Accordingly, the second order nonlinear optical effects are symmetry forbidden. The glasses in a conventional amorphous phase exhibit only a third order nonlinearity (THG) due to NLO four-wave mixing processes [4-6]. Inversion symmetry may however, be broken due to optical poling, as was confirmed on commercial silica glasses in several earlier studies, see e.g. Refs. [7,8]. In the corresponding NLO studies, a spatially periodic static electric field is induced due to the interference of fundamental ($\omega$) and doubled frequency ($2\omega$) laser beams; as a result there appears polarization grating and long-lived periodical second order susceptibility across the sample.

The OA in glass materials may also be induced, on the other hand, by intense polarized light of fundamental frequency $\omega$. As a particular example, the nonlinear Kerr effect is worth to be mentioned which results to NLO changes in the refractive index $n_i = n_0 + \delta n_i(I)$ induced

*Corresponding authors: E-mail addresses:* andriy.kityk@univie.ac.at (A.V.Kityk), jaroslaw.jedryka@o2.pl (J.Jedryka), vrksurya@rediffmail.com (V.R.Kumar).

by the intense continuous or pulsed laser light of intensity $I$. Nonlinear contribution $\delta n_i$ to the refractive index $n_0$ is defined by third order NLO susceptibilities' tensor components $\chi^{(3)}_{ijkl}$, i.e. $\delta n_i \propto \chi^{(3)}_{ijjj}(\omega)E^2_{0j} \propto I$, where $E_{0j}$ is the amplitude of the electric field [9]. In amorphous glasses or liquids the number of independent tensor components is reduced to two: $\chi^{(3)}_{1111} = \chi^{(3)}_{1221} + 2\chi^{(3)}_{1122}$ [10]. Taking into account that $\chi^{(3)}_{2211} = \chi^{(3)}_{1122}$ one arrives with inequality: $|\chi^{(3)}_{1111}| \neq |\chi^{(3)}_{2211}|$, i.e. the polarized pumping light ($E \parallel X$) induces uniaxial optical anisotropy ($n_1 \neq n_2 = n_3$) likewise the optical birefringence ($\Delta n = n_1 - n_2 \propto (\chi^{(3)}_{1111} - \chi^{(3)}_{2211})E^2_{01}$) observed in the direction of light propagation ($q \perp E$).

The Kerr effect depends on $E^2_{01}$ and is characterized by fast response; the changes in the refractive index in the optical materials may be caused when they are subjected to intense laser radiation. On the other hand, in the photorefractive effect, the change in the refractive index is a slow process and the changes can be preserved even after the light beam is switched off; however, such effects can be faded by intense uniform illumination [3]. The photorefractive effect usually occurs due to the inhomogeneous variation in the refractive index ($n$) causing defocusing and scattering of light. The origin of the photorefractive effect is based on the carrier transport mechanism. The slow photorefractive response is explained, therefore, by a slow diffusion of carriers. Tichomirov and Elliot [11] have observed a strong, optically anisotropic metastable photorefractive effect in a chalcogenide $As_2S_3$ glass and have demonstrated recording of one beam polarized self-induced hologram in this material. Characteristic feature of their experiment is, the optical anisotropy of the photoinduced grating in the glass by the polarized laser light.

Most of the studies on NLO refraction likewise photorefraction effects in isotropic materials are confined to silicate glass systems because of their broad optical applications. Among potential photorefractive materials, alkali sulfo-phosphate glasses are of specific interest. In such materials, the $SO_4^{2-}$ ions diffuse into phosphate glass network to a larger extent. Nevertheless, if $P_2O_5$ glass network contains metaphosphate groups, such diffusion is low and there may be a formation of thiophosphate $(SPO_7)^{3-}$ structural units in the material [12]. To illustrate it in different way, depending on the ratio of $P_2O_5$ and alkali sulphate groups, a sort of variations in linkages and de-linkages between $SO_4$–$PO_4$ units does occur in the glass matrix. Indeed, the P–O–S bonding in this amorphous system has been evidenced in NMR studies [13]. The exchange of interactions among sulphate and $PO_4$ units paves the way for the diffusion of alkali ions into the glass matrix.

Magnesium oxide is added to the glass matrix in order to make the sulfo-phosphate glasses more corrosion resistant. This modifying oxide is characterized by polyhedrons with oxygens of tetrahedral phosphate units which increase the aqueous durability of sulfo-phosphate glass [14,15].

The glasses with alkali/alkaline thiophosphate clusters and varied concentration of transition metal ions (TMI) are expected exhibiting high NLO and photorefractive efficiency when exposed to intense laser radiation [16,17]. Among different transition metal oxides containing glasses, NiO mixed glasses find potential applications in telecommunications since they exhibit strong luminescence in a broad spectral range with large gain [18,19]. The transition metal ions typically exist in different oxidation states in glass systems. Among them, the divalent nickel ions are highly stable in glasses and preferably reside in $O_h$ positions with large crystal field energy [20]. Because of such steady valency of $Ni^{2+}$ ions, NLO and photorefractive features of NiO mixed glasses are expected to be fascinating. In fact, a large number of recent works on different amorphous systems containing $Ni^{2+}$ ions are available in literature [21–23]. In addition to octahedral sites, a certain part of $Ni^{2+}$ ions may occupy tetrahedral ($T_d$) positions in glass network [24,25]. The ratio of ions which occupy $O_h$ and $T_d$ sites substantially depends both on the glass composition and the NiO content. The NiO doping in the glass host is expected to enhance its optical anisotropy when irradiated with high energetic fs laser.

Electrical and acoustic studies of $Li_2SO_4$–MgO–$P_2O_5$ glass system have been reported in our recently published works [26–28]. The free space trapping volume as a function of NiO content in the glass has been evaluated by positron annihilation (PAL) technique [29]. These studies evidently suggested that a portion of $Ni^{2+}$ ions in $O_h$ sites act as modifiers, increases with a rising NiO content. It induces, thereby, a structural chaos in the glass network and paves a way for the trapping of more cavities ultimately causing an increase of conductivity in the sulfo-phosphate glasses. Such varied magnitude of depolymerization in the glass network appears to be favorable for the photorefractive anisotropy induced by powerful femtosecond laser radiation.

In this work, we report NLO studies of $Li_2SO_4$–MgO–$P_2O_5$:NiO glasses. A combination of fs laser as a pumping light source and a high-accuracy polarimetry set-up with low energetic probing laser, is used to investigate the light induced birefringence (LIB) in these glass materials. The LIB in NiO-doped $Li_2SO_4$–MgO–$P_2O_5$ glasses exhibits slow relaxation character spreading over a few seconds evidently suggesting on anisotropic photorefractive effect besides fast Kerr effect. The obtained results are analyzed in the light of other recently reported experimental results and explained by the glass polymerization mechanisms. Recently, we have reported this type of studies in CuO doped glass system [17].

## 2. Experimental

The following mentioned composition of the glass (in mol%) is selected for this investigation: $20Li_2SO_4$–$20MgO$–$(60-x)P_2O_5$:$xNiO$ ($x$ = 0.2, 0.4, 0.6, 0.8 and 1.0) and the samples are named as $N_2$, $N_4$, $N_6$, $N_8$ and $N_{10}$ as per the quantity of NiO, respectively. The appropriate amounts of analar grade reagents of $Li_2SO_4$, $MgCO_3$, $P_2O_5$ and NiO were mixed thoroughly and melted in silica crucibles in the range of temperature 1000–1050 °C for about 1/2 h and were subsequently annealed from 350 °C. Other particulars of procedures adopted for the fabrication of the glasses and the methods used for their characterization, likewise the details of electrical conductivity, acoustic and positron annihilation measurements were reported in the Refs. [17,27–29].

Fig.1 presents a sketch of the experimental set-up used for LIB measurements. The set-up consists of a High Q-2 fs pulsed Yb-based laser as the pumping light source. The details of this laser beam are as follows: wavelength 1045 ± 5 nm, pulse duration ~250 fs with repetition rate 63 ± 0.6 MHz, FWHM 4–8 nm, peak radiant energy 40 nJ, ~2.5× $10^5$ W (in the duration of the pulse), The set-up further consists of a polarimetry arm with the continuous-wave (CW) of low power He–Ne laser ($\lambda$ = 633 nm, $P$ = 10 mW) used as the probing beam. The polarimetry part of the set-up contains the photo-elastic modulator (PEM-90, Hinds Instr.) inserted between crossed Glan-Thompson polarizer (P) and analyser (A). Modulated retardation ($\delta_m = \delta_0 \cos(\Omega t)$, $\Omega/2\pi$ = 50 kHz, $\delta_0$ = 0.383$\lambda$) results to a modulated light intensity, sensed by PD which is then analyzed simultaneously by a pair of lock-in amplifiers (SR-830), extracting amplitudes of first ($I_\Omega$) and second ($I_{2\Omega}$) harmonics of the modulated light. Measured optical retardation $\delta_x$, is then defined by the ratio of the measured harmonic amplitudes, $\delta_x = \tan^{-1}[I_\Omega J_2(\delta_0)/(I_{2\Omega} J_1(\delta_0)]$, where $J_1(\delta_0)$ and $J_2(\delta_0)$ are the Bessel function values ($J_2(0.383\lambda)/J_1(0.383\lambda) \approx 0.83$). The acquired data were transferred to PC via General Purpose Interface Bus (GPIB also known as IEEE-488 bus) that can be used to transfer data for real-time saving process. The accuracy of polarimetry measurements was ~5× $10^{-3}$ deg. which is equal to the precision in the birefringence measurements (~$10^{-8}$ for samples of thickness ~1 mm). For comparison, the precision of the refractive index measurements performing with most accurate interferometric set-ups does not exceed $10^{-5}$. Accordingly, the modulation polarimetry is able to provide the most accurate characterization of the anisotropic NLO and photorefractive effects, also weak ones being not accessible for the classical interferometry.

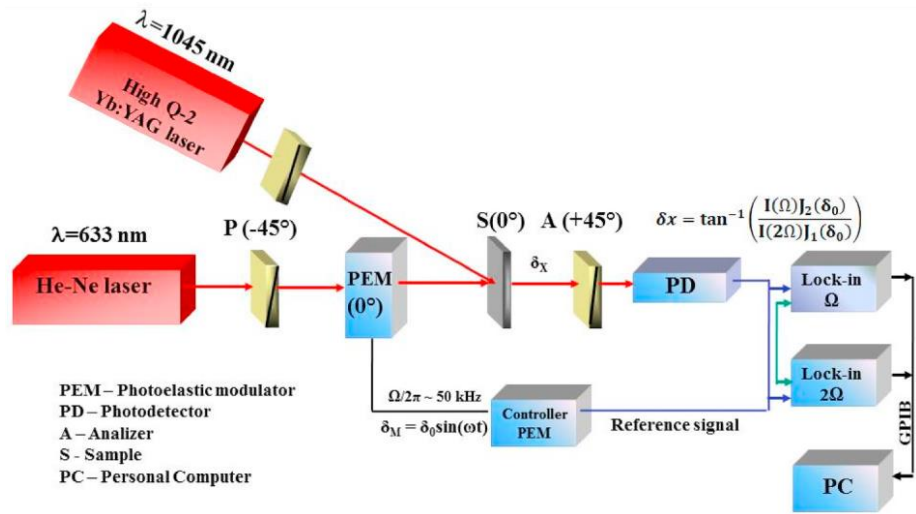

**Fig. 1.** Sketch of the experimental set-up used for measuring of photo induced birefringence (PIB) in the titled glass.

### 3. Results and discussion

Taking into account that the light induced birefringence (LIB) in NiO-doped glasses exhibit evident temporal behavior under intense light illumination. It has been recorded for each glass sample at regular intervals of time when the pump laser alternatingly switches between OFF/ON states. Fig.2 shows the temporal behavior of optical retardation (OR) for glass sample $N_8$ as an example. When the pumping laser light is switched ON or OFF, one observes first an abrupt, almost jump-like variations in the optical retardation which is accompanied then by its long-time exponentially saturation behavior spreading over a dozen of seconds. Our attention is concentrated on the light-induced optical anisotropy ignoring residual weak strain birefringence ($\sim 10^{-6}$) with locally random character and resulting apparently from the glass technology. Presented in Fig.3(a)-(e), the temporal circles of the LIB, being re-evaluated from measured retardation [$\delta(\Delta n) = \lambda \delta_x/(360^o d)$, $d$ is the sample thickness], provide most proper quantitative characterization of the light induced optical absorption excluding, particularly, the thickness factor.

The polarized pulsed laser radiation clearly induced the optical birefringence (OB) which may have different origin taking into account different time scales realized in its temporal behavior. As has been pointed out above, the glasses, representing macroscopically centrosymmetric media, reveal nonlinear Kerr effect [22]. When the pumping light is polarized, it may lead to the induced OA, i.e. the optical birefringence indeed demonstrated in our experiments. The Kerr effect is of electronic origin with extremely fast response. Accordingly, it may be attributed to the jump-like changes of the birefringence observed only right in the moments when pumping laser light is switched ON or OFF. Further slow temporal relaxation of the optical birefringence, clearly suggests the presence of some other mechanism of NLO, i.e., anisotropic photorefraction. Apparently, such slow relaxation may be ascribed to a varying degree of structural disorder in the glass network due to the change in the content of NiO. One should note that a certain photorefractive anisotropy has been revealed also in undoped samples; the samples have exhibited insignificant photorefractive birefringence Δn ($< 0.5 \times 10^{-6}$). The contribution of the Kerr effect to the light-induced birefringence cannot be precisely characterized. Our rough evaluations suggest, however, that its contribution to the overall changes of LIB should not exceed 5-10%, i.e., the light-induced anisotropy in NiO-doped glasses is strongly dominated by the photorefractive contribution.

Our earlier measurements (optical absorption and other spectroscopic studies [27–29]) suggested that the concentration of nickel ions, which reside in octahedral sites, increases as the content of NiO is increased up to 0.8 mol%. In the same range of NiO concentrations, the electrical conductivity demonstrated an enhancing trend with the content of NiO (Fig. 4(a) [29]), likewise, the free volume space entrenched in the glass network, as estimated by PAL spectroscopy ($^{22}$Na isotope, 0.1MBq), see Fig. 4(b) [29]. The ultrasonic velocity of longitudinal waves ($f = 4$ MHz) exhibits, in contrast, a decreasing trend with the NiO content (Fig. 4(c), [28]). Summarizing the results of these studies one may conclude that as the quantity of nickel oxide is increased up to 0.8 mol%, increases the degree of depolymerization of the glass network. Accordingly, it results in an increase of the photorefractive birefringence (see Fig. 3(f)) due to the low phonon losses as has been observed. The NiO content of 0.8 mol% may be considered here as optimal in the sense that at such concentration a major portion of nickel ions occupy octahedral sites and act as modifiers, with the internal de-augmentation of the glass network by disrupting P–O–S bonds [26,29]. In other words, the glass samples doped with such concentration of NiO reveal the largest photorefractive anisotropy. Such type of materials seemed to be therefore suitable for optical systems like optical circulators, interleaves, brightness enhancers in illuminating systems etc.

The decrease in the photorefractive anisotropy, as being observed in doped glass samples of higher NiO content (>0.8 mol%), suggests that tetrahedral sites become to be more preferable for location of nickel ions. In a structural aspect it corresponds to the opposite concentration

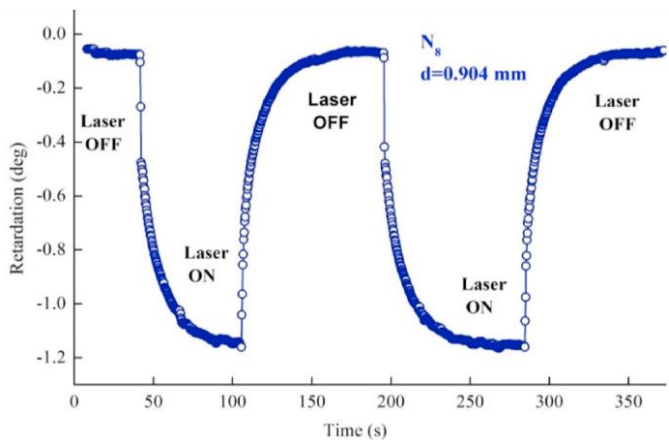

**Fig. 2.** Optical retardation vs time recorded at different temporal circles for the sample $N_8$ when the pumping laser light alternately switches between OFF and ON states.

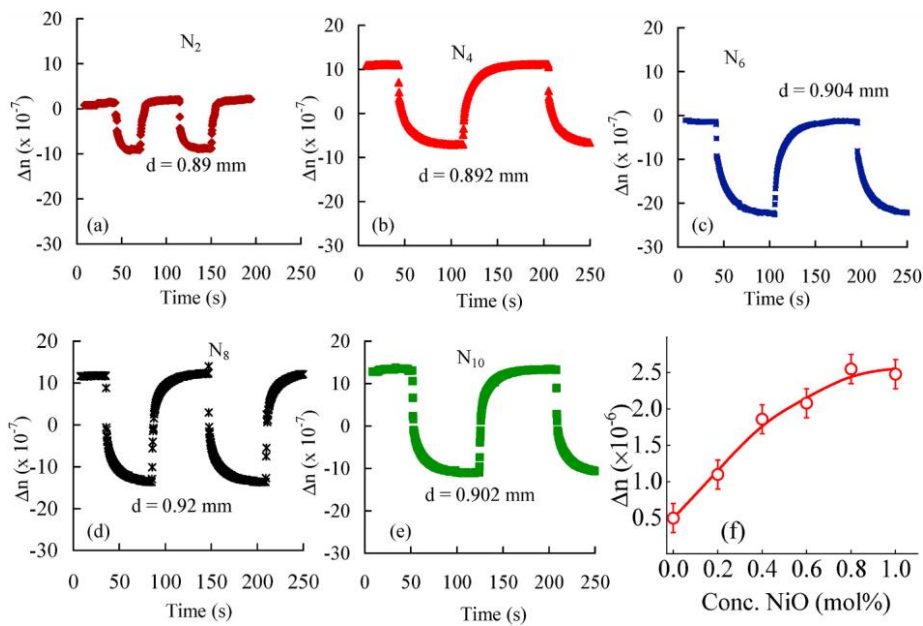

**Fig. 3.** Sections (a)–(e): Light-induced changes of optical birefringence in the studied glasses under alternately switching pumping laser light. Fig. 3(f) shows the change in light induced birefringence (LIB) with the NiO concentration.

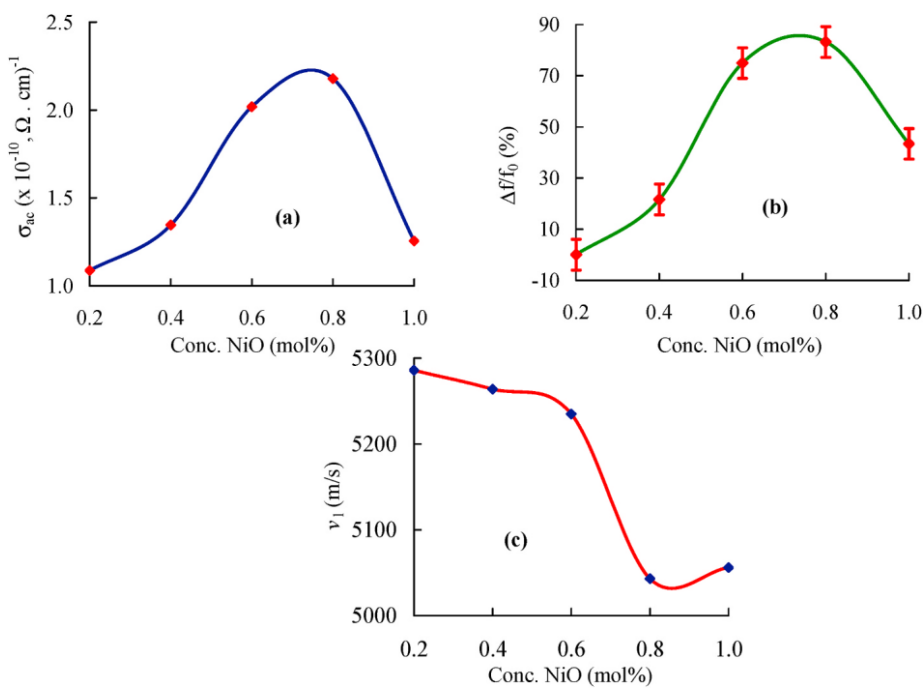

**Fig. 4.** (a) The plots of (a) a. c. conductivity (measured at 100 °C and 300 Hz) (b) free volume fraction trapped in the glass matrix (measured by PAL decay profiles recorded with 0.1 MBq $^{22}$Na isotope) and (c) longitudinal velocity of sound vs NiO concentration [27–29].

trend related, particularly, with an increase of the glass network polymerization.

### 4. Conclusions

Li$_2$SO$_4$–MgO–P$_2$O$_5$ glasses mixed with varied contents of nickel oxide were fabricated. Initial characterization by means of spectroscopic, electrical, PALS and ultrasonic studies, mentioned in detail in our earlier works, clearly suggested a rising degree of structural de-fragmentation in alkali sulfo-phosphate glass upon increase of nickel oxide concentration up to 0.8 mol%. It may be attributed to predominant occupancy of Oh sites by Ni ions in that act as modifiers and defragment PO$_4$ and SO$_4$ structural groups from P–O–S glass network. The LIB in NiO-doped Li$_2$SO$_4$–MgO–P$_2$O$_5$ glasses exhibits slow relaxation behavior. It is spread over a dozen of seconds evidently indicating on anisotropic photorefractive effect besides fast, jump-like Kerr response noticed instantly when pumping laser light is switched ON/OFF. The photorefractive birefringence increases with the NiO up to 0.8 mol%. Such behavior is ascribed to the rising degree of depolymerization of the glass network due to the increased number of Ni$^{2+}$ ions that occupy $O_h$ positions. Further increase of NiO content ( > 0.8 mol%) causes, however, a certain decrease of the photorefractive birefringence. We explain this by a rising phonon losses due to increasing portion of the nickel ions that occupy tetrahedral sites. In an aspect of the structural modification, it corresponds to the opposite trend, i.e. an increase of the glass network polymerization. Taking together, the doped glasses with the NiO content of about 0.8 mol% may be considered as optimal in the sense of photorefractive efficiency. Relevant samples that have exhibited largest magnitudes of the photorefractive birefringence appear to be favorable for potential applications.

**CRediT authorship contribution statement**

**A. Siva Sesha Reddy:** Conceptualization, Methodology, Investigation. **A.V. Kityk:** Methodology, Data curation, Formal analysis, Software, Writing – original draft. **J. Jedryka:** Methodology, Data curation, Formal analysis, Software, Writing – original draft. **N. Purnachand:** Methodology, Data curation, Formal analysis, Software, Writing – original draft, Conceptualization, Methodology, Investigation. **P. Rakus:** Methodology, Data curation, Formal analysis, Software, Writing – original draft. **A. Wojciechowski:** Methodology, Data curation, Formal analysis, Software, Writing – original draft. **A.S. Andrushchak:** Methodology, Data curation, Formal analysis, Software, Writing – original draft. **V. Ravi Kumar:** Conceptualization, Methodology, Investigation, Supervision, Writing – original draft, Writing – review & editing. **N. Veeraiah:** Supervision, Writing – original draft, Writing – review & editing.

**Declaration of competing interest**

The authors declare that they have no known competing financial interests or personal relationships that could have appeared to influence the work reported in this paper.

**Acknowledgement**

The presented results are part of a project that has received funding from the European Union Horizon 2020 research and innovation programme under the Marie Sklodowska-Curie grant agreement no. 778156. Support from resources for science in years 2018–2022 granted for the realization of international co-financed project Nr W13/H2020/2018 (Dec. MNiSW 3871/H2020/2018/2) is also acknowledged. N. Veeraioah, wishes to thank UGC, New Delhi, for sanctioning BSR Faculty fellowship to carry out this work.